# Diffusion of $^{210}$Pb and $^{210}$Po in nylon

P. Adhikari,[*] M. G. Boulay, R. Crampton, M. Perry, and D. Sinclair
*Department of Physics, Carleton University, Ottawa, Ontario K1S 5B6, Canada*

Radon and its progeny constitute a major source of background in rare-event physics experiments, such as those searching for dark matter, neutrinos, and neutrinoless double $\beta$ decay, due to their origin as unavoidable decay products of natural uranium. In particular, $^{222}$Rn and its long-lived daughter $^{210}$Pb can diffuse from detector material surfaces, resulting in sustained background contributions. To investigate this process, a system was developed using a controlled radon source, a vacuum chamber with a high electric field, and a thin nylon-6 film to enable deposition of radon progeny onto the film surface. Nylon-6 was selected for the initial measurement given its history in low-background experiments. We intend to systematically study diffusion in various polymers in the future. Our setup allowed for controlled study of the diffusion behavior of $^{210}$Pb and its daughter $^{210}$Po under varying humidity conditions. Our results show that both $^{210}$Pb and $^{210}$Po diffuse significantly in nylon under high relative humidity, which can potentially lead to internal contamination and increased background in low-background detectors. The diffusivity of $^{210}$Pb was found to be lower than $1.14 \times 10^{-15}$ cm$^2$/s at 40% relative humidity (RH) and to be $(4.03 \pm 1.01) \times 10^{-13}$ cm$^2$/s at 95% RH. The diffusivity of $^{210}$Po at 95% RH was measured to be $(3.94 \pm 0.98) \times 10^{-13}$ cm$^2$/s. These findings underscore the importance of controlling environmental humidity and material exposure to radon in the design of ultralow-background experiments.

## I. INTRODUCTION

Experiments that focus on the search for rare events, such as those targeting dark matter and neutrinoless double $\beta$ decay, demonstrate a high sensitivity to radioactivity emanating from both the detector materials and the surrounding environment. These detectors are situated deep underground to shield the background induced due to the cosmic rays on the surface of Earth [1–4]. Despite stringent controls such as material assays, careful selection, and handling in a clean environment, detectors continue to face a persistent $\alpha/\beta$ background source. This interference arises from the daughter products of the $^{222}$Rn decay chain, which originates from the decay of $^{238}$U [5]. It is known that radon can diffuse into materials, where it may decay to form the long-lived progeny $^{210}$Pb. Extreme care is needed to achieve surface and bulk cleanliness levels of $^{210}$Pb and $^{210}$Po required by sensitive experiments, and even so diffusion of these radon daughters may lead to contamination levels beyond those required. In this study, we collected $^{210}$Pb from decay of $^{222}$Rn onto the surface of a thin nylon film and investigated its diffusion into the bulk of the film. Nylon-6, commonly employed in low-background experiments as a container for detector materials and storage, is particularly susceptible to $^{210}$Pb diffusion, posing a critical background source for such experiments.

$^{210}$Pb possesses a long half-life of 22.2 yr, contributing to a consistent background level throughout the typical lifetime of an experiment if it contaminates the detector. $^{210}$Pb decays to $^{210}$Bi through $\beta$ emission as shown in Eq. (1) [6] with a relatively low Q-value of 63.5 keV, but the $^{210}$Bi decays through $\beta$ emission to $^{210}$Po with a Q-value of 1162.1 keV, and $^{210}$Po emits the α particle of energy 5304 keV before becoming to stable $^{206}$Pb; this $\alpha$ is a significant source of background in many rare-event search experiments. Due to the relatively short half-life of $^{210}$Po, the primary concern in long-term experiments is the diffusion of $^{210}$Pb with its long half-life,

$$^{210}\mathrm{Pb} \xrightarrow[\beta(63.5\ \mathrm{keV})]{22.2\ \mathrm{yr}} {}^{210}\mathrm{Bi} \xrightarrow[\beta(1162.1\ \mathrm{keV})]{5.01\ \mathrm{days}} {}^{210}\mathrm{Po} \xrightarrow[\alpha(5304\ \mathrm{keV})]{138.37\ \mathrm{days}} {}^{206}\mathrm{Pb}. \tag{1}$$

The radioisotopes deposited on the surface diffuse into the film according to the diffusion equation,

$$\frac{\partial C(x,t)}{\partial t} = D\frac{\partial^2 C(x,t)}{\partial x^2}. \tag{2}$$

[*]Contact author: pushparajadhikari@cunet.carleton.ca

The thin film membrane is modeled as a semi-infinite slab, and since the particle diffuses along its thickness, we consider it as a one-dimensional diffusion phenomenon. There are several initial and boundary conditions in our particular case of source deposition and diffusion.

(i) Initial condition:
A deposited source at the surface

$$C(x,0) = Q\delta(x) \tag{3}$$

where Q is the deposited amount of the source.

(ii) Boundary conditions:
Semi-infinite slab

$$C(x,t) \to 0 \quad \text{as } x \to \infty. \tag{4}$$

No continuous flux at $x = 0$,

$$\frac{\partial C(0,t)}{\partial x} = 0. \tag{5}$$

Equation (2) can be solved to satisfy the initial condition of a surface-deposited source, together with the boundary condition appropriate for a semi-infinite slab. Representing the nylon surface with zero continuous flux at the $^{210}$Pb deposition interface, the solution to Eq. (2) yields the following:

$$C(x,t) = \frac{Q}{\sqrt{4\pi Dt}} \exp\left(-\frac{x^2}{4Dt}\right), \tag{6}$$

where

(i) Q is the deposited amount of the source;
(ii) D is the diffusion coefficient;
(iii) $x \geq 0$ is the spatial coordinate;
(iv) $t > 0$ is time.

In our *ex situ* measurements, elaborated in Secs. II and III, we observed the diffusion of $^{210}$Pb and $^{210}$Po, and the results are consistent with the pattern shown in Fig. 1. Additionally, we calculated the diffusivity of these isotopes in nylon. We chose nylon and investigated the dependence on humidity since it is known that diffusion of radon is significant in nylon and has a strong dependence on humidity [7]. Also, we plan to systematically measure various polymers used in low-background experiments under different ambient conditions.

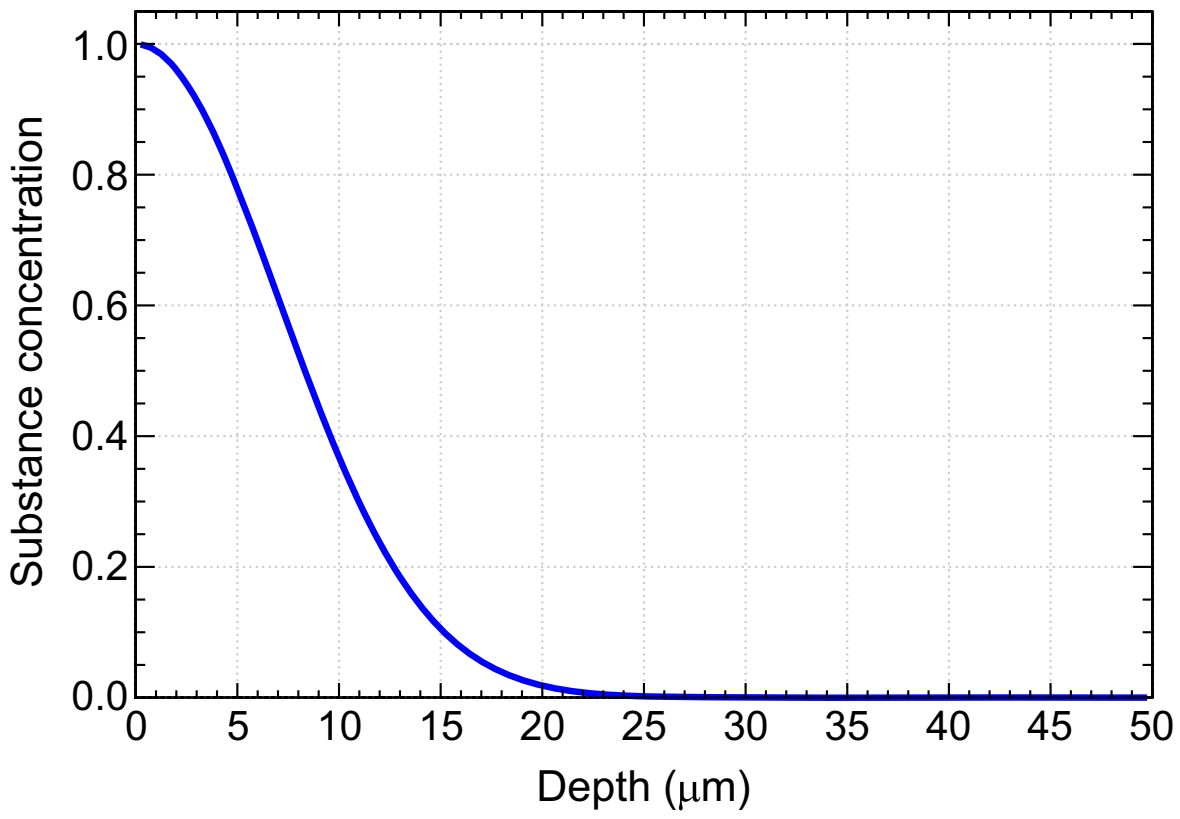

FIG. 1. Concentration of substance within a thin membrane due to diffusion.

## II. EXPERIMENTAL SETUP

The experiment is divided into two distinct sets of measurements. First, there is the deposition of $^{210}$Pb on a nylon film within a deposition chamber. Second, the $\alpha$ activities and energy distributions of $^{210}$Po are monitored using an $\alpha$ spectrometer, Ortec Alpha Duo. The $\alpha$ counter uses the ULTRA-AS series detectors, which have a thin (500 Å) contact that is ion implanted into the silicon surface. The sample size, vacuum pump requirement, electronics, and software information of the $\alpha$ counter are described in [8]. The subsequent sections provide a detailed discussion of the $^{210}$Pb deposition process and its corresponding measurements.

$^{222}$Rn is an inert noble gas produced in the radioactive decay of $^{238}$U, subsequently undergoing further decay to form the long-lived $^{210}$Pb. The decay series is illustrated as follows [9]:

$$^{222}\text{Rn} \xrightarrow[\alpha(5.49\text{ MeV})]{3.82\text{ days}} {}^{218}\text{Po} \xrightarrow[\alpha(6.00\text{ MeV})]{3.05\text{ min}} {}^{214}\text{Pb} \xrightarrow[\beta(730\text{ keV,max})]{26.8\text{ min}} {}^{214}\text{Bi} \xrightarrow[\beta(3.27\text{ MeV,max})]{19.7\text{ min}} {}^{214}\text{Po} \xrightarrow[\alpha(7.69\text{ MeV})]{0.16\text{ ms}} {}^{210}\text{Pb}. \tag{7}$$

The nylon sample has a diameter of 2.5 cm; however, the surface area exposed to the source corresponds to the diameter of the acrylic tube, which is 1 cm. The thickness of the sample is 50 μm. To deposit $^{210}$Pb onto the surface of this nylon film, we exposed the film to a concentrated source, model 2000A radon source from Pylon Electronics with an activity 25 kBq [10] in a small chamber with a controlled electrostatic field. The field intensity and gas pressure within the tube are optimized to maximize the deposition of charged daughter nuclides onto the film. The increased surface activity from deposited daughters exceeds the activity of those formed from radon that has diffused into the film. In this way, we end up with a distinctive signal from $^{210}$Po $\alpha$'s at the film surface and can search for diffusion of daughters in the film by observing their degraded energy spectra in the cases that are generated from regions below the surface.

The fundamental concept and design are illustrated in Fig. 2. In this setup, the radon source is connected to an acrylic tube, operating with nitrogen gas at approximately 200 mbar. The system features a pressure gauge to measure the nitrogen pressure inside the tube, capable of measuring pressures up to 1000 mbar below atmospheric pressure, a valve for pumping, and a nitrogen gas Dewar connection. $^{222}$Rn decays in the gas volume, producing $^{218}$Po. The positively charged Po ions drift under the applied electric field toward a foil mounted at one end of the tube. The tube is positioned within electrodes and supports made of printed circuit boards (PCBs), featuring a hole matching the outer diameter of the acrylic tube. Twenty PCBs, spaced 1-cm apart, are arranged and connected with 10 MΩ resistors to maintain a uniform electric field along the tube. A 130-MΩ resistor is connected between the final PCB to sample holder. As the tube approaches the foil, the electric field increases by approximately a factor of 10. A 3.5-kV high voltage is applied across the tube to generate the electric field, focusing the ions onto a spot on the foil at the tube's end. This arrangement allows enhancement of daughter deposition onto the surface of the foil in relation to the diffusion of radon into the bulk and its subsequent contamination with decay progeny. The efficiency of radon daughter deposition on the film will be discussed in Sec. III.

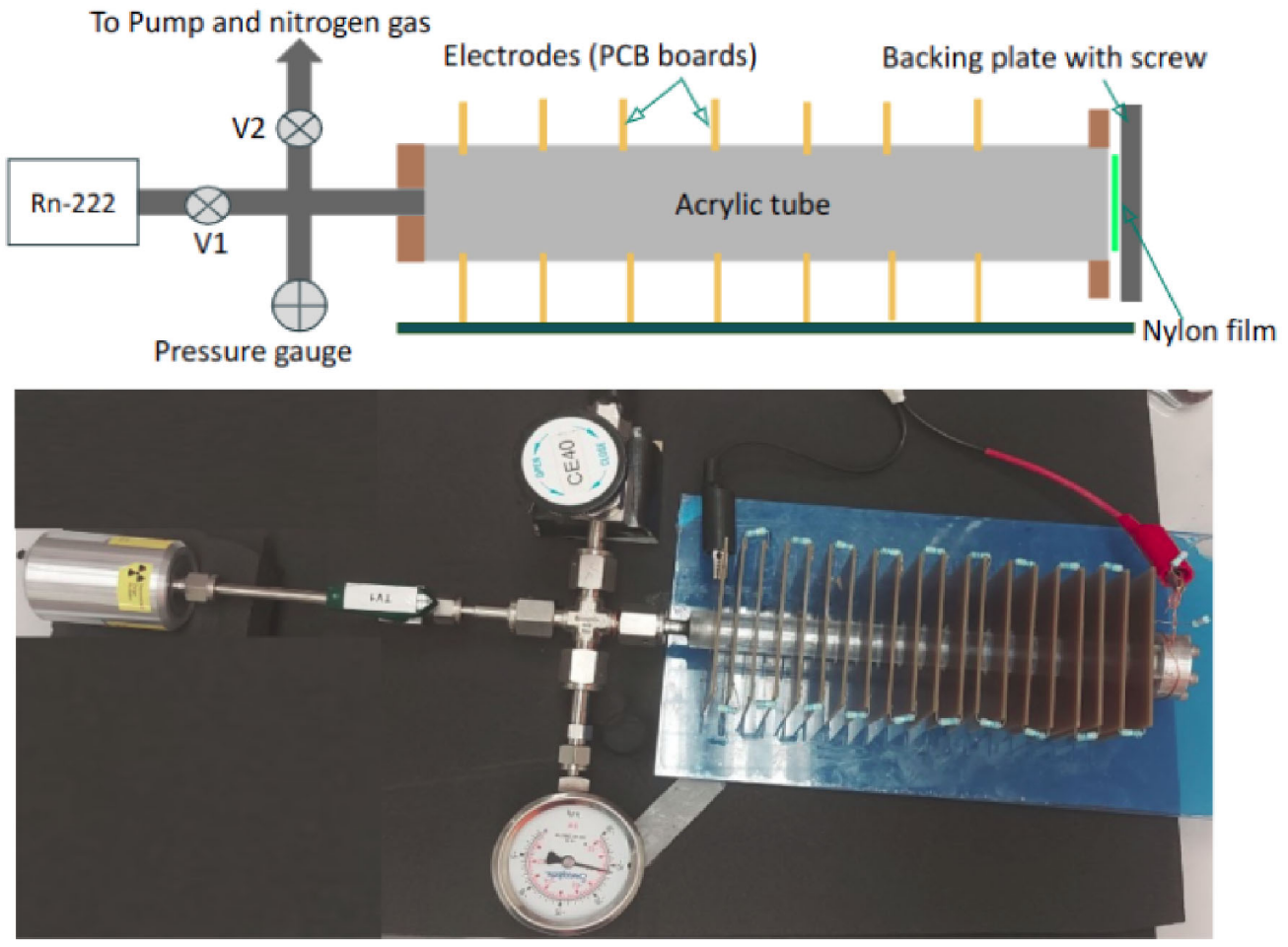

FIG. 2. Top: diagram of the radon daughter deposition chamber showing the components. Bottom: photo of the assembled system.

The samples are stored in a clean room of class 10000 considered in an ISO7 cleanroom at 40% relative humidity and can also be subjected to a humidity chamber set at 95% relative humidity to investigate the dependence of diffusion on humidity. Samples are retrieved from storage every week for 24 h and then placed on the $\alpha$ counter to measure $\alpha$ activities and energy distribution. This process is repeated over several months to observe the energy spectra and from those infer the diffusion pattern.

## III. ANALYSIS

### A. Calibration and efficiency

The $\alpha$ spectrometer efficiency and energy scale were calibrated according to the following prescription. The $\alpha$ spectrometer features ten slots positioned at varying distances from the detector for placing samples. Efficiency is dependent on the distance between the detector and the sample. A $^{241}$Am source emitting 5.486-MeV monoenergetic $\alpha$ particles was positioned at various locations to assess $\alpha$ detection efficiency. The identical source was employed for energy calibration. The $\alpha$ counter and a typical $^{241}$Am $\alpha$ peak are shown in Fig. 3. The analog-to-digital converter count is converted to energy in keV based on the $\alpha$ energy generated [11]. As the sample approaches the detector, there is a noticeable degradation in $\alpha$ energy resolution, yet the efficiency experiences an increase. We selected slot number 8, positioned 10 mm away from the detector, to observe the nylon sample. This location provides relatively good resolution, with approximately 15% detection efficiency.

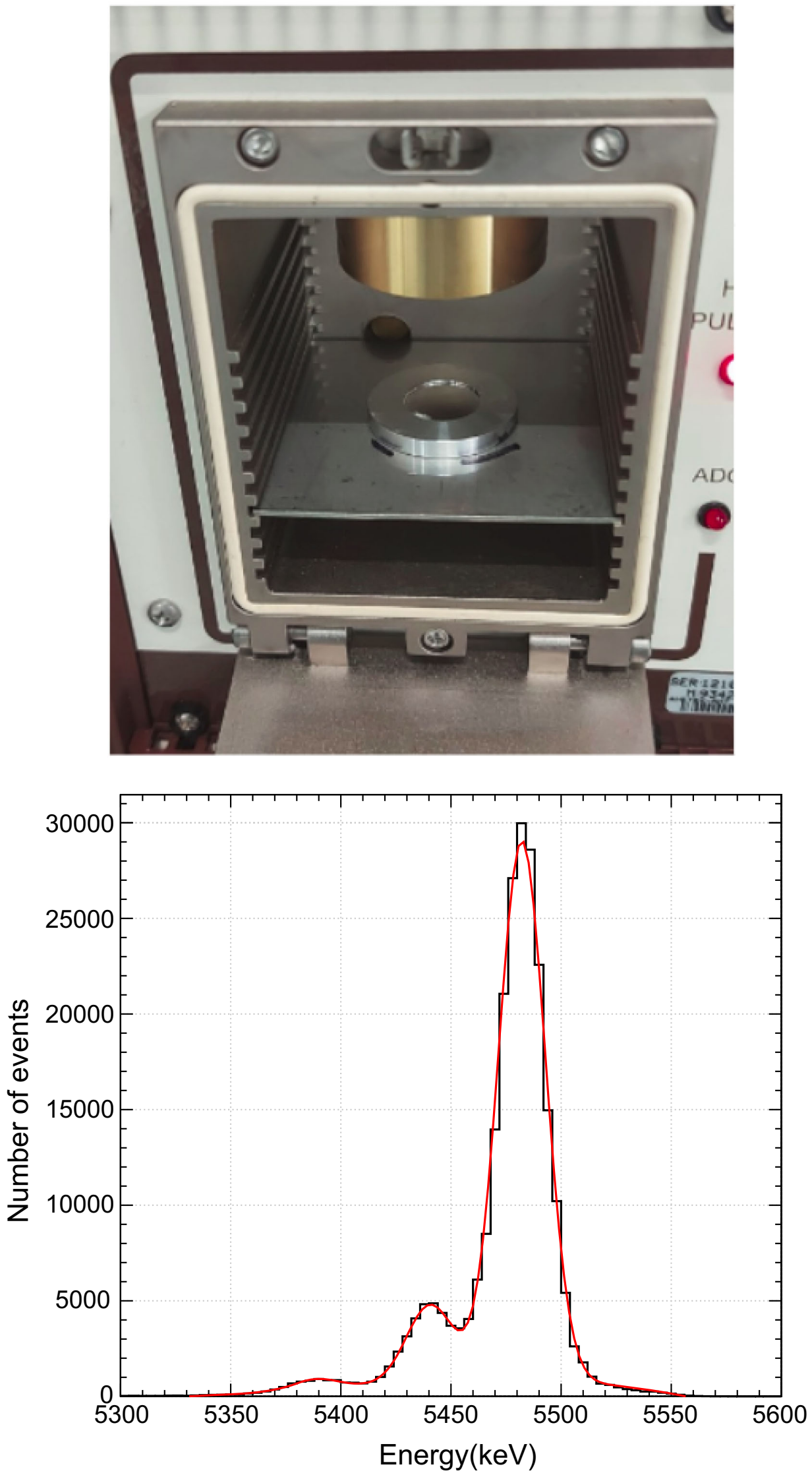

FIG. 3. Top: a photo of $\alpha$ spectrometer in which an $^{241}$Am source is placed for calibration and efficiency estimation. Bottom: an $\alpha$ energy spectrum from the $^{241}$Am source.

The collection efficiency of radon daughters from the source is not 100% due to several factors. One source of inefficiency arises from radon decay within the source. This inefficiency can be approximated by the ratio of the tube volume to the source volume where, "source volume" is the container with the uranium/radon source itself and the associated plumbing that connects it to the high voltage (HV) chamber, and "tube volume" is the volume within the HV chamber where daughters can be accelerated in the electric field. Additionally, some of the Po daughters may not carry a charge because of the neutralization of their charge in the tube or may be negatively charged. For example, the positively charged fraction of freshly formed $^{218}$Po is $87.3 \pm 1.3$ [12]. In the case of uncharged Po atoms, there is a second chance for collection when Po decays. However, for negative ions, there is a loss.

Furthermore, if Po decays occur in the foil, some daughters may recoil off the foil. In this scenario, they are likely to be charged, and the electric field in the setup can accelerate them toward the foil, mitigating some of the losses. These factors collectively contribute to the overall efficiency of radon daughter collection, and understanding and quantifying these inefficiencies are crucial for accurate interpretation of experimental results.

The efficiency of radon daughter deposition can be estimated by counting the number of $^{214}$Po $\alpha$ emissions observed on the film immediately after the sample is taken from the deposition chamber. $^{214}$Po is prevalent in the sample due to its short decay time, as illustrated in the top part of Fig. 4. By fitting the $\alpha$ rate observed in the sample, as shown in the bottom part of Fig. 4, one can determine the deposition rate of $^{214}$Po ions on the sample.

The fit provides the initial rate of $^{214}$Po, representing the deposition rate of radon daughters on the nylon sample, estimated at approximately 13 Hz. This rate suggests that 50% of the daughters deposited on the film originated from $^{222}$Rn within the HV tube.

Additionally, the fit provides the decay time of the $^{214}$Po peak. According to Eq. (7), the longest half-life among the first four $^{222}$Rn progeny is 26.8 min. Consequently, equilibrium between radon and its progeny is achieved within a few hours [13].

The rate of deposition provides the total number of $^{210}$Pb atoms deposited on the sample over the course of 15 days. Consequently, one can estimate the $\alpha$ emissions induced by $^{210}$Po from the sample using Eq. (8),

$$R = N \times 0.693/t_{\frac{1}{2}} \tag{8}$$

where R is the rate of events or activity of the source, N is the number of deposited nuclei, and $t_{\frac{1}{2}}$ is the half-life of the $^{210}$Pb. The saturated $\alpha$ rate is found to be approximately 10 Hz, which is consistent with the measurement shown in Fig. 5.

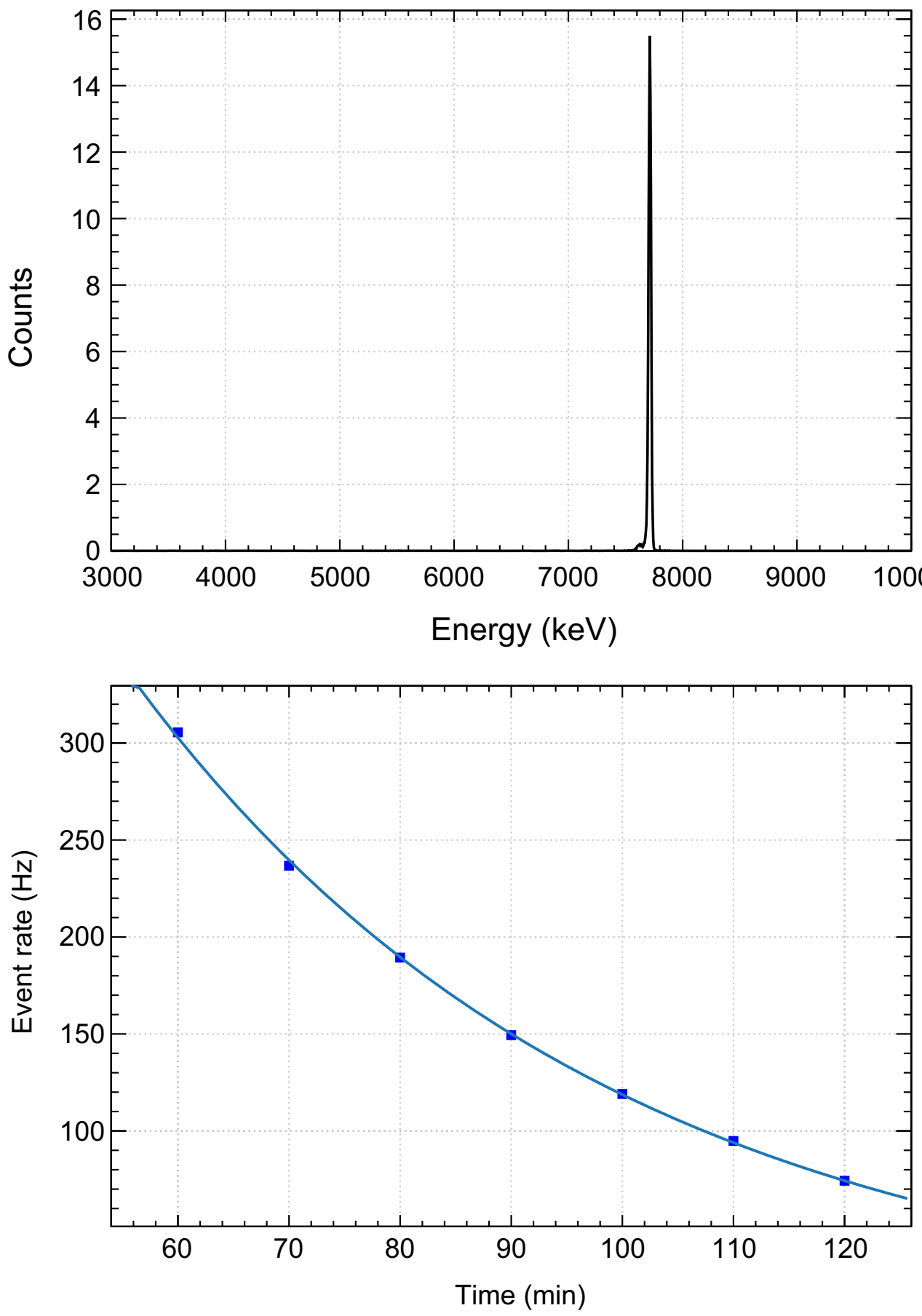

FIG. 4. Top: $^{214}$Po induced $\alpha$ energy from the sample within a few hours after removal from the deposition chamber. Bottom: the $^{214}$Po rate versus time. The $^{214}$Po events are the integration of 7.69-MeV peak.

### B. Sample measurement

The deposited $^{210}$Pb transforms into $^{210}$Po and $^{206}$Pb through $\alpha$ decay, as indicated in Eq. (1). If the $^{210}$Po contamination is initially zero at the time of $^{222}$Rn exposure, it will increase with a characteristic time of $\tau_{Po} = 199.64$ days until equilibrium is attained [14]. This growth can be described by

$$R_{\alpha} = A_0(1 - e^{-(t-t_0)/\tau_{Po}}) \tag{9}$$

where $R_{\alpha}$ is the rate of $^{210}$Po after time t, and $t_0$ is the time when the initial $^{210}$Pb contamination occurred. The increased rate of $^{210}$Po $\alpha$ and fitted with Eq. (9) is shown in Fig. 5.

The $^{210}$Pb-deposited nylon film was stored at room humidity for approximately 200 days and then placed inside a humidity chamber of control humidity 95% relative humidity (RH) for an additional 6 days to observe the impact of humidity. Our current setup has humidity chamber with controlled RH of 95%; we will follow up in a subsequent paper with measurements at various humidities.

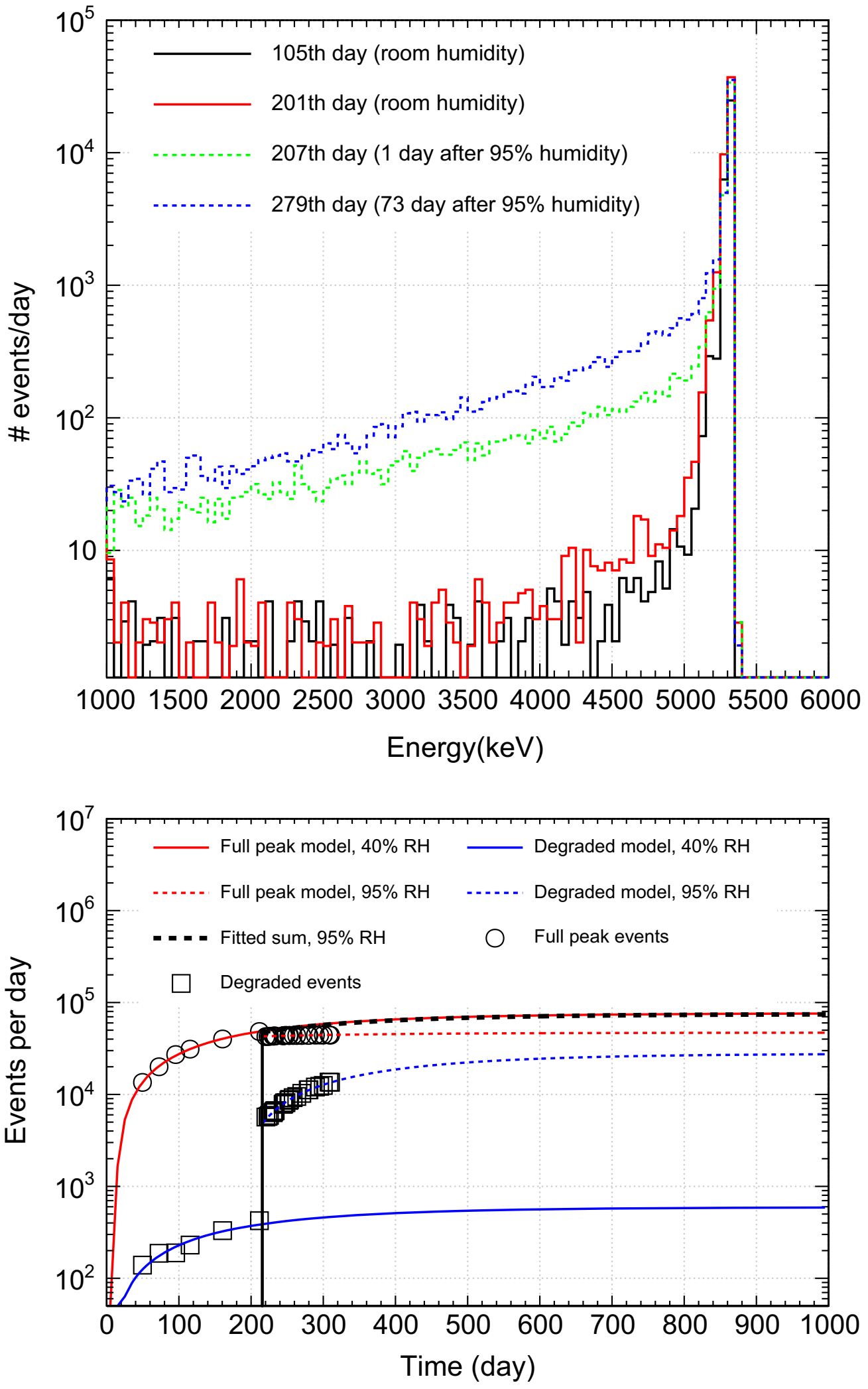

FIG. 5. The top part of the figure presents a comparison of $\alpha$ energy from the nylon sample on different days. Two lines, black and red, represent the condition before exposing the sample to humidity, while the other two lines depict the state after humidity contact. Notably, there is a substantial increase in low-energy events after humidity exposure, implying diffusion within the sample. In the bottom part, the events versus time distribution are illustrated, fitted with Eq. (10) to estimate the number of saturated events or total events due to $^{210}$Po activity on the sample. The two solid lines represent the condition before humidity exposure, indicating that degraded $\alpha$ events are minimal (about 1% of total events). In contrast, the dashed lines, depicting the situation after humidity contact, show a significant increase in degraded $\alpha$ events over time. This observation suggests a substantial diffusion of $^{210}$Po in the sample.

The $\alpha$ energy spectrum and the number of $\alpha$ events from the sample were closely monitored over a period of 300 days. The continuous growth of $^{210}$Po induced $\alpha$ emissions in the sample, as described by Eq. (9), and aligns with the radon daughter deposition rate shown in Fig. 5.

The full energy spectrum indicates under 40% relative humidity with minimal or no diffusion observed. Then, a significant amount of $^{210}$Po and $^{210}$Pb diffused onto the nylon layer after exposing the sample to 95% relative humidity for six days.

The sample was observed under two conditions: normal lab humidity (40% RH) and inside the humidity chamber at 95% RH. First, the sample, stored in an acrylic Petri dish in a clean room (40% RH) for 200 days, was subjected to data collection in the $\alpha$ counter every week. Approximately 1% of the $^{222}$Rn and its initial daughter products appear to have diffused onto the sample while it was inside the source chamber, as indicated by the initial measurement. A similar 1% fraction of degraded $\alpha$ events relative to the full $\alpha$ peak is also observed in both the 105- and 200-day datasets. We attribute this fraction to a combination of initial diffusion and diffusion occurring under 40% humidity. However, due to the limited statistics and insufficient power to distinguish the origin of these events, we report only an upper limit on the diffusion, which will be explained later. Also, comparison of $\alpha$ distributions shown in the top plot of Fig. 5 indicates significant increases in degraded $\alpha$ events after humidity encounter, indicating that the diffusion occurs in the nylon layer.

Events versus time plots were fitted using the following equation:

$$\mathrm{Fitval} = \mathrm{P}[0](1 - \mathrm{e}^{-(t-t_0)/\tau_{\mathrm{Po}}} - \mathrm{P}[1]) \qquad (10)$$

where P[0] gives the saturated $\alpha$ rate due to $^{210}$Po decay, P[1]*P[0] are the initial events present in the sample at the contamination time $t_0$, and $\tau_{\mathrm{Po}} = 199.64$ days is the characteristic time until equilibrium. The fitted curves offer insights into the dynamics of $^{210}$Po diffusion over time. In Fig. 5, bottom plot, two solid lines are the full peak $^{210}$Po events which are surface-induced events and the degraded $\alpha$ events (1.0–5.1 MeV) before contact with the source in high humidity. The two dashed lines, blue and red, are the energy-degraded events and full-energy events, respectively, after humidity exposure, and the black dashed line is total events after humidity. A 2% reduction in the total number of detected events was observed following humidity exposure. This decrease is likely attributable to event diffusion deeper into the sample, causing signals to fall below the energy detection threshold, as well as lateral diffusion away from the sample center, leading to reduced detector efficiency.

### C. $^{210}$Po and $^{210}$Pb diffusion

The sample measurements indicate signs of source diffusion from the nylon surface. Both radioisotopes, $^{210}$Po and $^{210}$Pb, present on the surface, may diffuse when exposed to high humidity. It is essential to distinguish between diffusion events of $^{210}$Po and $^{210}$Pb. As previously mentioned, detecting $\alpha$ events from $^{210}$Pb requires a long waiting period, whereas diffusion events of $^{210}$Po can be observed immediately after humidity exposure. Therefore, the data collected immediately after exposure, represented by the green line in the top panel of Fig. 5, primarily reflect $^{210}$Po diffusion.

During exposure in the humidity chamber, $^{210}$Po diffuses into the sample and continues to decay throughout the measurement interval due to its 138.37-day half-life. Consequently, the initial $^{210}$Po inventory decreases over time but still contributes noticeably to the measured spectrum, while the amount of $^{210}$Po generated from $^{210}$Pb diffusion increases as described by Eq. (9). To isolate the events produced solely by $^{210}$Pb diffusion, the number of $\alpha$ emissions attributable to the decay of the initial $^{210}$Po was calculated and subtracted. Figure 6 illustrates both the initial $^{210}$Po and its remaining fraction on the sample.

The diffused $^{210}$Po is decayed with time as the half-life of it is 138.37 days and at the same time diffused $^{210}$Pb increases following Eq. (9). To see the events due to $^{210}$Pb diffusion we need to subtract the observed events to the remaining $^{210}$Po in that particular time. So we have calculated the fraction of initial $^{210}$Po that remain on the sample and subtracted it with the total events on the sample. Figure 6 shows that initial $^{210}$Po and its remaining fraction on the sample.

### D. Simulation and diffusion estimation

We simulated a 5.3-MeV $\alpha$ particle within a 50-μm thick nylon sample. By applying position (thickness) cuts, we obtained energy spectra corresponding to both the surface and various depths within the nylon layer. These simulated spectra were then fitted to experimental data to estimate the $\alpha$ particle contribution from different surface depths.

The fitting process involved adjusting the spectrum coverage fraction of each depth as a free parameter, allowing us to determine the number of events originating from various thicknesses beneath the surface. The results of this fit are shown in Fig. 7 (top), where the data that were collected one day after the sample was exposed to 95% relative humidity. Figure 7 illustrates the dependence of diffusion on the material thickness. As the thickness increases, the propagation distance of the nuclide decreases exponentially. Owing to instrumental noise below 1 MeV, a 1-MeV cutoff is applied.

To estimate the diffusivity, we plotted and fitted the number of events as a function of surface depth using Eq. (6). This fitting provided the diffusivity of $^{210}$Po on the nylon surface. An additional parameter was included in the fitting to account for background events in the data. The resulting diffusivity of $^{210}$Po obtained from the fitting is $3.9 \times 10^{-5}$ $\mu\mathrm{m}^2/\mathrm{s}$ ($3.9 \times 10^{-13}$ $\mathrm{cm}^2/\mathrm{s}$) under the condition of 95% relative humidity.

Similarly, the $^{210}$Pb spectrum shown in Fig. 6 (bottom) is fitted using a free-floating parameter approach, and the number of events as a function of surface depth is analyzed using Eq. (6). Since multiple $^{210}$Pb spectra were collected on different dates, each spectrum was individually fitted, consistently yielding similar diffusivity values across all dates, despite variations in the number of events. An example of the $^{210}$Pb fitting process is shown in Fig. 8 (top and middle), while the fitted diffusivity results from different dates are presented in the bottom plot of Fig. 8.

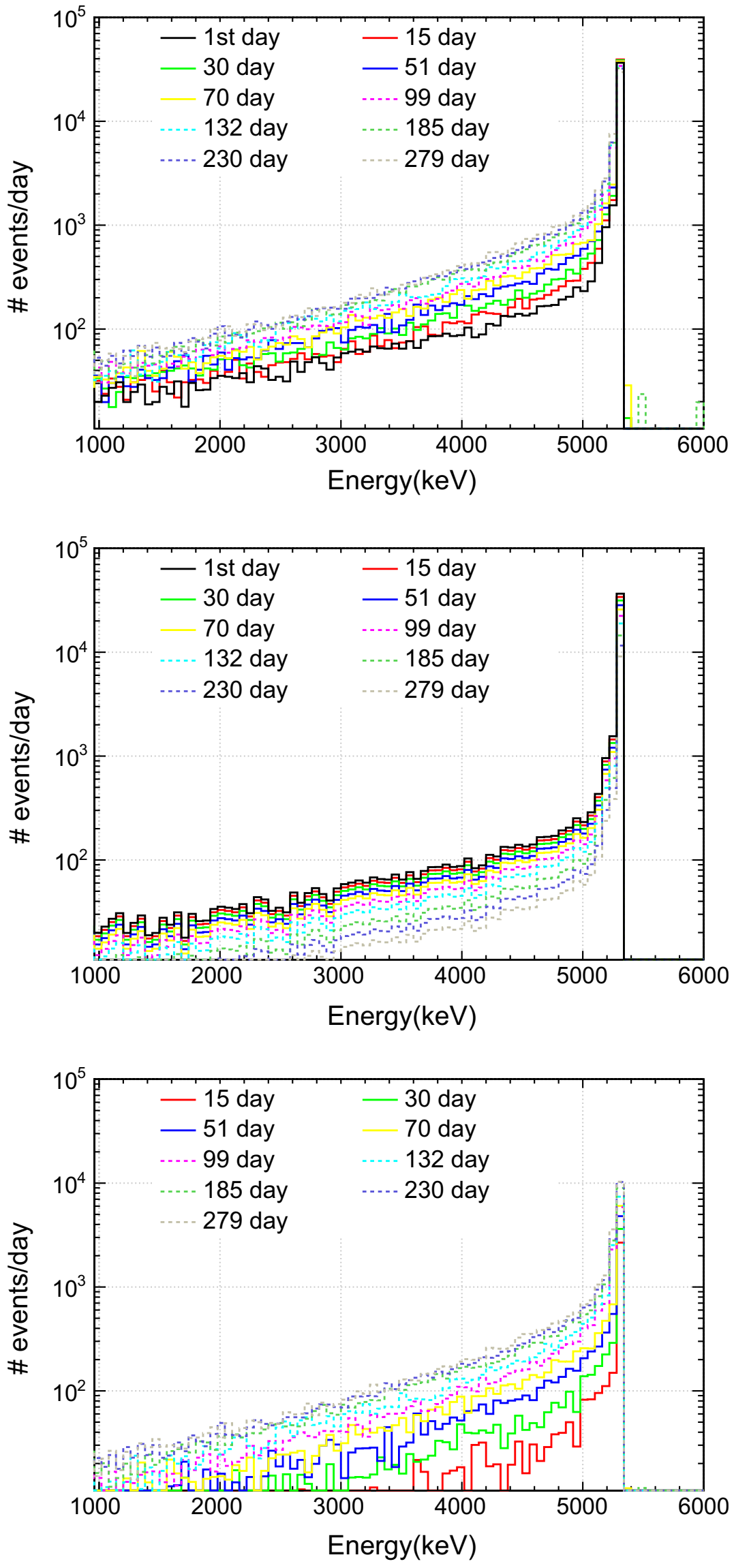

FIG. 6. Top: events observed after the sample was removed from the humidity chamber. Each spectrum represents 24 h of data, with the labels indicating the time elapsed since humidity exposure. The data from the first day primarily reflect $^{210}$Po diffusion, while the increasing number of low-energy events on later days indicates the diffusion of $^{210}$Pb. Middle: the black line represents 24-h data collected immediately after the nylon sheet sample was taken out of the humidity chamber, primarily capturing $^{210}$Po diffusion events. The spectra labeled with different days correspond to data after the decay of $^{210}$Po, which has a half-life of 138.37 days, relative to the first-day events. Bottom: events attributed to $^{210}$Pb diffusion while the sample was inside the humidity chamber. These plots are obtained by subtracting the contribution of remaining $^{210}$Po diffusion from the total observed events.

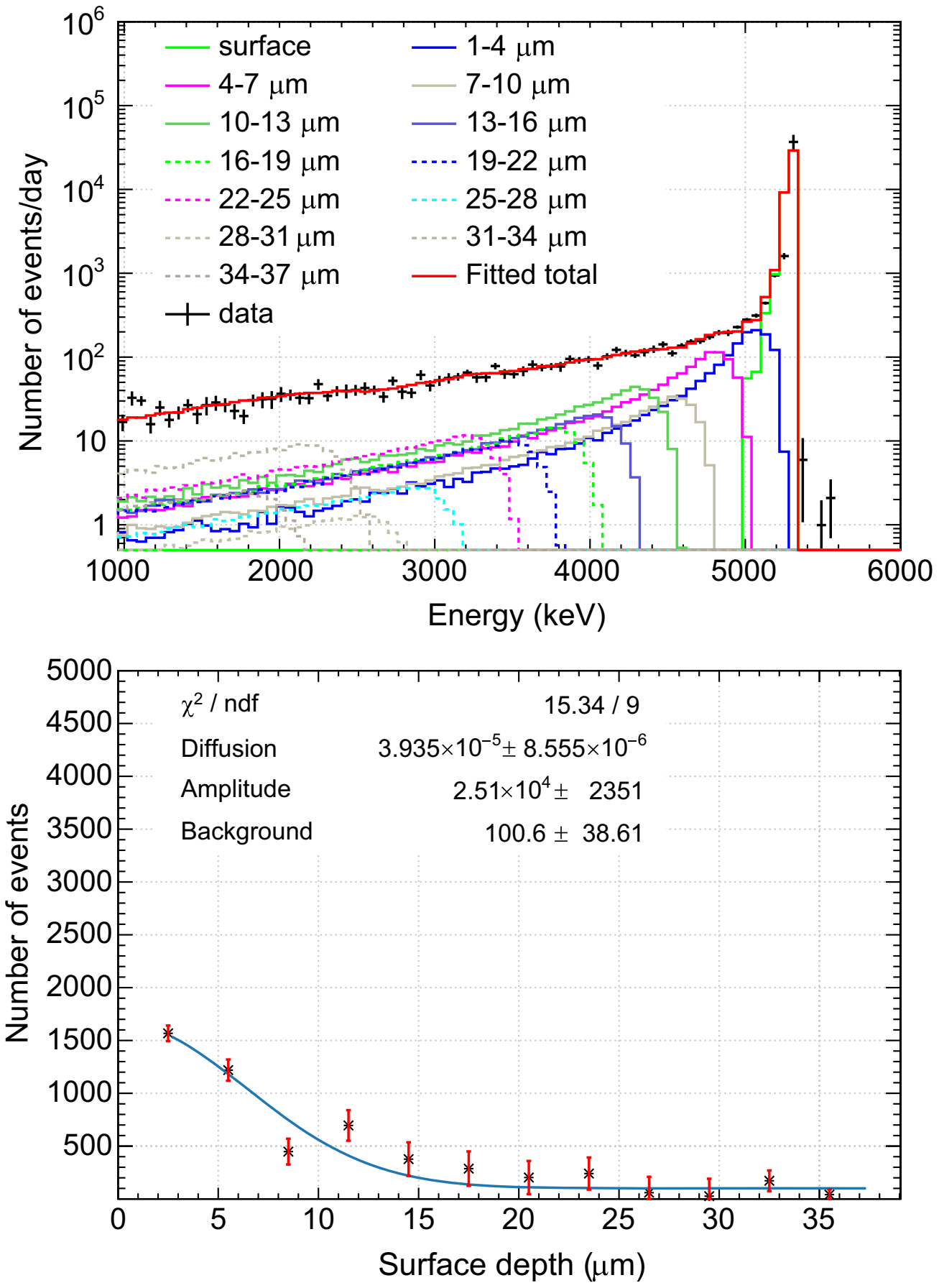

FIG. 7. Top: free-floating fitting of the $\alpha$ spectrum from different nylon thicknesses, based on data collected 1 day after the sample was exposed to 95% relative humidity for 6 days. Bottom: the number of events as a function of surface depth is fitted using Eq. (6), yielding a diffusivity of $3.9 \times 10^{-5}$ $\mu m^2/s$.

As the sample remained in room humidity (40% RH) for 200 days, there were minimal events observed in the degraded energy region. The fitted data indicate that the diffusivity is approximately 1000 times lower than that of the sample exposed to 95% RH. However, due to lower statistics and higher uncertainty in the fit value, we can only provide an upper limit for the diffusivity.

### E. Systematic uncertainties

Systematic effects impacting this measurement must be carefully evaluated and incorporated into the analysis. The primary sources of systematic uncertainty arise from the data acquisition system and the fitting procedures, which are integral to the measurement. The leading contributions to the systematic uncertainty in the calculated diffusion rates are summarized below.

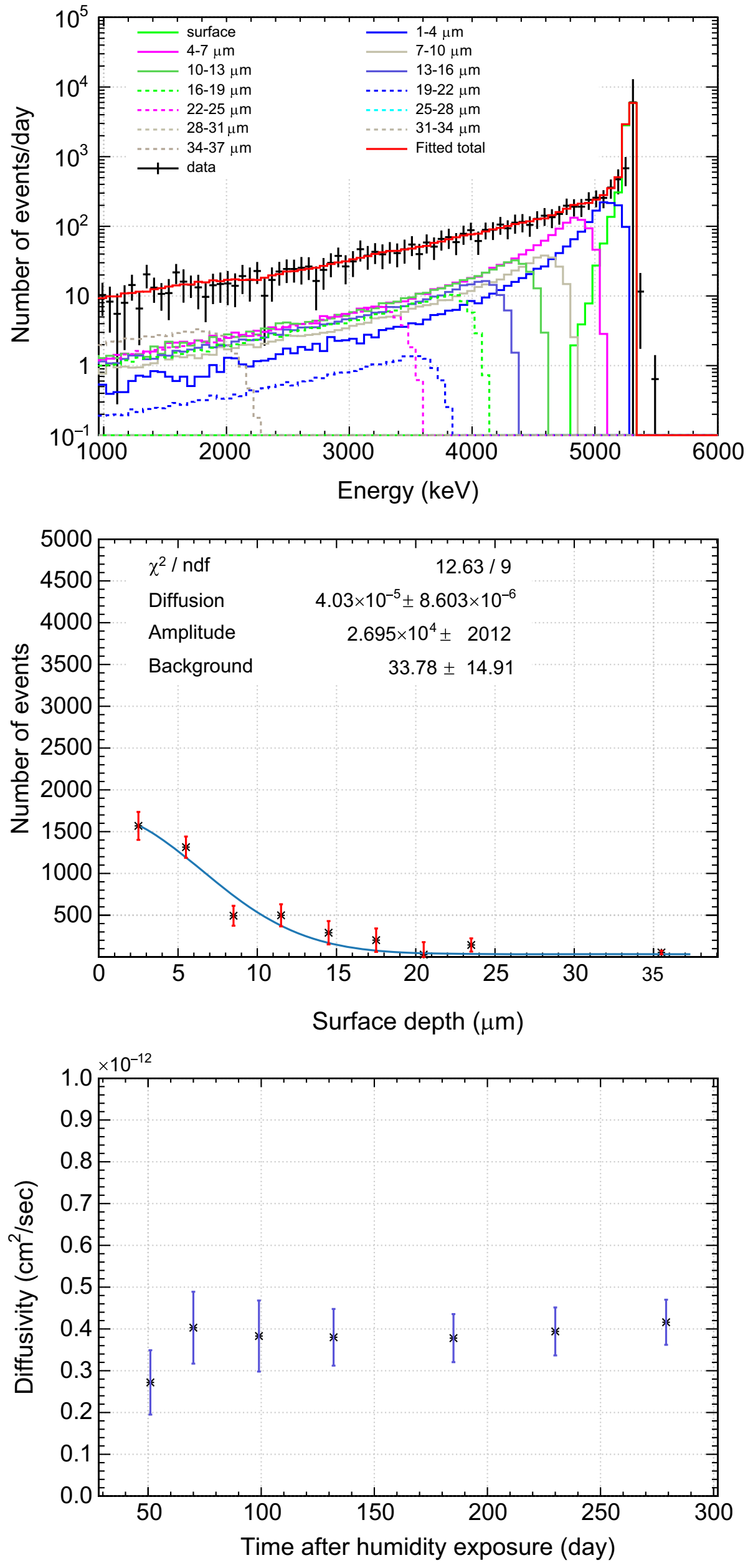

FIG. 8. Top: free-floating fitting of the $\alpha$ spectrum from various nylon thicknesses, using the $^{210}$Pb-only spectrum. This spectrum is obtained by subtracting the remaining $^{210}$Po diffusion events from the data collected 70 days after exposure to 95% relative humidity. Middle: the number of events as a function of surface depth is fitted using Eq. (6). Bottom: fitted diffusivity values obtained from $^{210}$Pb data collected on different dates after humidity exposure.

(i) Position of sample in the $\alpha$ counter: A systematic uncertainty arises from the positioning of the sample within the Ortec counter during data acquisition. The nylon film was mounted on a stainless steel base plate inside the counter crate, maintaining a fixed vertical distance between the sample and the detector. Although care was taken to consistently place the sample at the same location, slight variations in the X–Y position can affect the $\alpha$ particle collection efficiency and event rates.

To quantify this effect, measurements were performed using a calibrated $^{241}$Am $\alpha$ source placed at various positions on the base plate. The maximum deviation in lateral positioning was found to be 4 mm, corresponding to a variation in detection efficiency of up to 13%. This positional uncertainty represents a significant systematic effect and must be considered in the interpretation of the $\alpha$ measurement data.

(ii) Decay of the isotope: Systematic uncertainties also arise from the decay of isotopes during the periods when the sample was placed inside the radon collection chamber and the humidity chamber. $\alpha$ counting and spectral fitting began only after the sample was removed from these chambers. Consequently, any decay of $^{210}$Po occurring during these inactive periods may not have been fully recorded, potentially leading to an underestimation of the total activity. Specifically, the sample remained in the radon collection chamber for 15 days prior to the start of a 200-day counting period and later spent an additional 6 days in the humidity chamber. These uncounted decay intervals introduce a systematic uncertainty that must be accounted for in the interpretation of the measured $^{210}$Po activity.

To account for the potential loss of $^{210}$Po decays while the sample was inside the chambers, the data were fitted using three different initial times ($t_0$) in Eq. (10): the first corresponding to the day the sample was placed inside the chamber, the second (nominal) corresponding to the midpoint of the chamber exposure period, and the third to the day the sample was removed. This range of $t_0$ values provides a systematic treatment of the uncertainty associated with unrecorded decays during the chamber exposure. The resulting variation in fitted parameters corresponds to an estimated uncertainty of approximately 6%. This systematic uncertainty should be taken into account when interpreting the final results.

(iii) Energy scale: The $^{210}$Po $\alpha$ peak was fitted with a Gaussian function to monitor potential energy scale shifts over time. While the mean energy of the peak remained largely stable throughout the measurement period, a gradual increase of approximately 0.3% in the peak centroid was observed over 300 days. This variation is attributed to potential long-term drift in the energy calibration and is conservatively assigned as a systematic uncertainty in the energy scale.

(iv) Fit uncertainty: Fitting the data for samples with varying thicknesses introduces a spread of up to 20% in the extracted fit parameters. This variation is treated as a systematic uncertainty associated with the fitting procedure and must be included in the overall uncertainty budget.

Considering these sources of systematics, the total systematic error was estimated using standard deviation variations in error propagation. Addressing these uncertainty, the $^{210}$Pb and $^{210}$Po diffusivity observed in 95% relative humidity are $(4.03 \pm 1.01) \times 10^{-13}$ and $(3.94 \pm 0.98) \times 10^{-13}$ cm$^2$/s, respectively.

## IV. SUMMARY AND OUTLOOK

Radon daughter diffusion can be a significant background source with implications for low-background experiments. This work investigates the diffusion of $^{210}$Po and $^{210}$Pb in the nylon films. Experiments sensitive enough to directly detect $^{210}$Po decays may observe background events far exceeding their expected levels and in unexpected regions. An even more concerning scenario involves experiments unable to directly observe $^{210}$Po decays, where $^{210}$Po could introduce unnoticed background interference.

The diffusivity of $^{210}$Po and $^{210}$Pb in a nylon film was estimated under 40% and 95% relative humidity conditions. In the case of 40% humidity, the number of events in the quenched $\alpha$ energy region, explaining the diffusion, is suppressed by background events. Therefore, an upper limit was calculated based on the background events, resulting in an upper limit of diffusivity in 40% humidity of $< 1.14 \times 10^{-15}$ cm$^2$/s.

The $^{210}$Pb diffusivity observed in 95% relative humidity is $(4.03 \pm 1.01) \times 10^{-13}$ cm$^2$/s. The significant increase of diffusivity in high humidity conditions may be attributed to the polar structure and the mechanical degradation of nylon in elevated temperature and humidity levels, as correlated with the material's diffusion behavior [15,16]. The diffusivity of $^{210}$Po at 95% relative humidity is found to be $(3.94 \pm 0.98) \times 10^{-13}$ cm$^2$/s.

We have investigated the diffusion of $^{210}$Po and $^{210}$Pb at humidity levels of 40% and 95%, observing an approximately 1000-fold increase in diffusivity between these levels. Currently, we are examining diffusivity at additional humidity values within this range to establish a humidity versus diffusion trend. This may help identify a critical humidity level where diffusion may significantly increase.

## ACKNOWLEDGMENTS

We thank the Natural Sciences and Engineering Research Council of Canada, the Canadian Foundation for Innovation (CFI), the Ontario Ministry of Research and Innovation (MRI), Carleton University, the Canada First Research Excellence Fund, and the Arthur B. McDonald Canadian Astroparticle Research Institute.

## DATA AVAILABILITY

The data that support the findings of this article are not publicly available. The data are available from the authors upon reasonable request.